\documentclass[12pt]{iopart}
\usepackage{graphicx}
\begin{document}

%\title{Activation experiments on the $^3$He($\alpha,\gamma$)$^7$Be reaction including the LUNA results}
\title[Comparison of the LUNA $^3$He($\alpha,\gamma$)$^7$Be activation results with ...]{Comparison of the LUNA $^3$He($\alpha,\gamma$)$^7$Be activation results with earlier measurements and model calculations}

\author{Gy~Gy\"urky$^1$, D~Bemmerer$^{2,3}$, F~Confortola$^4$, H~Costantini$^4$, A~Formicola$^5$, R~Bonetti$^6$, C~Broggini$^2$, P~Corvisiero$^4$, Z~Elekes$^1$, Zs~F\"ul\"op$^1$, G~Gervino$^7$, A~Guglielmetti$^6$, C~Gustavino$^5$, G~Imbriani$^8$, M~Junker$^5$, M~Laubenstein$^5$, A~Lemut$^4$, B~Limata$^8$, V~Lozza$^2$, M~Marta$^{3,6}$, R~Menegazzo$^2$, P~Prati$^4$, V~Roca$^8$, C~Rolfs$^9$, C~Rossi Alvarez$^2$, E~Somorjai$^1$, O~Straniero$^{10}$, F~Strieder$^9$, F~Terrasi$^{11}$, H P~Trautvetter$^9$}

\address{$^1$ Institute of Nuclear Research (ATOMKI), Debrecen, Hungary}
\address{$^2$ Istituto Nazionale di Fisica Nucleare (INFN), Sezione di Padova, Padova, Italy}
\address{$^3$ Forschungszentrum Dresden-Rossendorf, Dresden, Germany}
\address{$^4$ Universit\`a di Genova and INFN Sezione di Genova, Genova, Italy}
\address{$^5$ INFN, Laboratori Nazionali del Gran Sasso (LNGS), Assergi (AQ), Italy}
\address{$^6$ Istituto di Fisica Generale Applicata, Universit\`a di Milano and INFN Sezione di Milano, Italy}
\address{$^7$ Dipartimento di Fisica Sperimentale, Universit\`a di Torino and INFN Sezione di Torino, Torino, Italy}
\address{$^8$ Dipartimento di Scienze Fisiche, Universit\`a di Napoli ''Federico II'', and INFN Sezione di Napoli, Napoli, Italy}
\address{$^9$ Institut f\"ur Experimentalphysik III, Ruhr-Universit\"at Bochum, Bochum, Germany}
\address{$^{10}$ Osservatorio Astronomico di Collurania, Teramo, and INFN Sezione di Napoli, Napoli, Italy}
\address{$^{11}$ Seconda Universit\`a di Napoli, Caserta, and INFN Sezione di Napoli, Napoli, Italy}
\ead{gyurky@atomki.hu}

\begin{abstract}
Recently, the LUNA collaboration has carried out a high precision measurement on the $^3$He($\alpha,\gamma$)$^7$Be reaction cross section with both activation and on-line $\gamma$-detection methods at unprecedented low energies. In this paper the results obtained with the activation method are summarized. The results are compared with previous activation experiments and the zero energy extrapolated astrophysical $S$ factor is determined using different theoretical models. 
\end{abstract}

\pacs{25.55.-e, 26.20.+f, 26.35.+c, 26.65.+t}
\maketitle

\section{Introduction}

The precise knowledge of the $^3$He($\alpha,\gamma$)$^7$Be reaction rate is essential in two distinct fields of nuclear astrophysics: in the hydrogen-burning of the Sun (and similar main sequence stars) and in the big-bang nucleosynthesis. In the pp-chain of solar hydrogen burning the $^3$He($\alpha,\gamma$)$^7$Be reaction is the staring point of the 2nd and 3rd branches of the chain from where the high energy $^7$Be and $^8$B neutrinos are originated. Owing to the fast technical development of neutrino detectors, the study of the solar neutrinos has entered a high precision era. The flux of the $^8$B neutrinos has already been measured by the SNO and SuperKamiokande
neutrino detectors \cite{aharmim05,hosaka06} with a precision of 3.5\,\% and similar precision is foreseen for the $^7$Be neutrino flux in the near future \cite{alimonti02}. As the solar neutrino puzzle has been solved by the experimental observation of neutrino oscillation \cite{ahmad01,eguchi03}, the precise knowledge of the solar neutrino fluxes allows a more detailed test of the prediction of the Standard Solar Model (SSM). The SSM uses a number of input parameters to predict the solar neutrino fluxes. These input parameters include on one hand some solar properties (like the luminosity, radiative opacity, diffusion, elemental composition) and on the other hand the rates of nuclear reactions involved in the pp-chain. The uncertainty in the input parameters translates directly into uncertainties in the neutrino flux prediction. Therefore, the reduction of the input parameter uncertainties is required to carry out a more stringent test of SSM. From the nuclear physics parameters the rate of the $^3$He($\alpha,\gamma$)$^7$Be reaction has the highest uncertainty. Its 9\% error \cite{adelberger98} at solar energies (which was reduced to about 5\% by a recent work \cite{narasingh04}) contributes 8\% \cite{bahcall04} to the uncertainty in the predicted fluxes for solar $^7$Be and $^8$B neutrinos.

From the very precise measurement of the anisotropy of the cosmic microwave background radiation by the WMAP satellite, the barion-to-photon ratio in our Universe is known with high accuracy \cite{spergel03}. Since this ratio is the only free parameter in big-bang nucleosynthesis calculations, the abundances of the primordial elements (D, $^3$He, $^4$He, $^7$Li) can be calculated unambiguously and the results can be compared with astronomical observations. While there is a perfect agreement between the observations and calculations for the D abundance and reasonably good agreement for $^4$He, the predicted abundance of $^7$Li is a factor 2 to 3 higher than the observed one. At the measured barion-to-photon ratio $^7$Li is produced almost exclusively by the $^3$He($\alpha,\gamma$)$^7$Be reaction followed by the $\beta$-decay of $^7$Be. Therefore, the abundance of primordial $^7$Li depends on the rate of this reaction. The precise knowledge of this reaction rate is the necessary basis of the search for possible solutions to the  $^7$Li problem.

\section{Experimental investigation of $^3$He($\alpha,\gamma$)$^7$Be}

The rate of the $^3$He($\alpha,\gamma$)$^7$Be can be determined from the reaction cross section at energies relevant for the astrophysical site in question. Hydrogen burning in the core of the Sun takes place at roughly 15$\times$10$^6$K temperature corresponding to a relevant energy range of 10 to 35 keV\footnote{When not stated otherwise, energy always means center-of-mass energy throughout this paper} for the $^3$He($\alpha,\gamma$)$^7$Be reaction. Owing to the higher temperatures involved, the relevant energy range for the big-bang nucleosynthesis spans from 160 to 380 keV. At these deep sub-Coulomb energies the reaction cross section has very low values from a few hundreds of nanobarn at the highest big bang energies down to the attobarn range for the pp chain energies. While direct measurement of the cross section is possible at big-bang energies, the cross section at solar energies can only be calculated by extrapolating the high energy data towards low energies using theoretical considerations. The extrapolation is normally done with the help of the astrophysical $S$ factor which does not contain the trivial energy dependence of the Coulomb barrier penetration and which has a finite value at zero energy. Therefore, the $S$ factor is usually extrapolated to zero energy instead of a (temperature dependent) finite energy of astrophysical relevance. Since the $^3$He($\alpha,\gamma$)$^7$Be reaction at low energies proceeds through the direct capture process, different theoretical approaches lead to very similar energy dependence of the $S$ factor, though small deviations can be found. The only free parameter in the extrapolations for a certain theoretical model is the absolute magnitude of the $S$ factor. This parameter can be obtained by normalizing the theoretical energy dependence to measured values at energies where data are available. Therefore, it is important to measure the cross section of $^3$He($\alpha,\gamma$)$^7$Be with high precision and at energies preferably relevant directly to the big-bang nucleosynthesis and low enough that the extrapolation to zero energy could be done with low uncertainty despite the small differences in various theoretical energy dependences.

The mechanism of the $^3$He($\alpha,\gamma$)$^7$Be provides two possibilities for the measurement of the cross section via $\gamma$-spectroscopy. The reaction has a Q value of 1.586 MeV and at low energies the direct capture can lead to the ground state and first excited state in $^7$Be with the emission of two single $\gamma$-lines with energies of E$_{c.m.}$\,+\,Q and E$_{c.m.}$\,+\,Q\,-\,429\,keV. In the latter case the first excited state of $^7$Be decays to the ground state with a 429\,keV $\gamma$-emission. The detection of these three $\gamma$-lines provides one possibility for the cross section measurement (prompt-$\gamma$ or on-line method). The other method for the cross section determination is the measurement of the produced $^7$Be activity. By the measurement of the 478\,keV $\gamma$-radiation following the electron-capture decay of $^7$Be can be used to determine the total number of fusion reactions (activation method). 

Until the 1980's the $^3$He($\alpha,\gamma$)$^7$Be cross section has been measured seven times \cite{holmgren59,parker63,nagatani69,krawinkel82,osborne82,alexander84,hilgemeier88} with the on-line method and three times with activation \cite{osborne82,robertson83,volk83}. The experiments have been carried out in a wide range of energies and using very different experimental techniques. The lowest energy reached with the on-line method was $E$\,=\,107\,keV. This means that on-line experiments cover the energy relevant for the big-bang nucleosynthesis but are rather far from the solar energy. With activation, however, the lowest measured energy was 897\,keV \cite{robertson83}, far above the astrophysical energies.

Each data sets have been used to extrapolate the $S$ factor to zero energy. The extrapolations led to an ambiguous result when considering the on-line and activation experiments separately. Activation experiments obtained higher $S$ factor values than the on-line measurements resulting in a discrepancy of about 13\% (roughly 2.5\,$\sigma$ deviation) at zero energy. For the solution of this discrepancy several possible hidden experimental errors have been proposed. These include the not precisely known angular distribution of the prompt-$\gamma$ radiation, a possible weak monopole transition in the reaction hidden for the on-line experiments, or the presence of parasitic reactions leading to $^7$Be production in the activation case. %However, none of these reasons could have been proven to be the cause of the discrepancy. 

As more and more precise data became available for both the primordial $^7$Li abundance and solar neutrino fluxes, and as the precision of other reaction rate determinations relevant for BBN and Solar hydrogen burning has been dramatically increased, the need for a more precise $^3$He($\alpha,\gamma$)$^7$Be reaction rate measurement became pressing. 

\section{''Modern'' measurements of $^3$He($\alpha,\gamma$)$^7$Be}

Realizing the need for new, high precision cross section data, B. S. Nara Singh \textit{et al.} has performed an activation experiment for the $^3$He($\alpha,\gamma$)$^7$Be reaction \cite{narasingh04}. The cross section was measured at four energy points between 420 and 950 keV. Although these results were the lowest energy activation points ever measured, they were still above the astrophysically relevant energies. Moreover, an activation-only experiment -- whatever precise it is -- is not able to examine the discrepancy between the on-line and activation data. Therefore, an experiment measuring the $^3$He($\alpha,\gamma$)$^7$Be $S$ factor simultaneously with both techniques is highly needed. Such a measurement should be carried out with high precision and at energies as low as possible in order to have a reliable zero energy extrapolation of the $S$ factor.

\begin{table}
\caption{\label{tab:results} Results of the LUNA activation experiments}
\begin{tabular}{lcccc}
\hline
Phase of the experiment & E$_{c.m.}$ & $S$ factor & Stat. error & Syst. error \\
\hline
combined & 92.9 & 0.534 & 0.016 & 0.017 \\
activation only & 105.6 & 0.516 & 0.027 & 0.016 \\
combined & 105.7 & 0.493 & 0.015 & 0.015 \\
activation only & 126.5 & 0.514 & 0.010 & 0.015 \\
activation only & 147.7 & 0.499 & 0.008 & 0.015 \\
activation only & 168.9 & 0.482 & 0.010 & 0.015 \\
combined & 169.5 & 0.507 & 0.010 & 0.015 \\
\hline
\end{tabular}
\end{table}

Recently, the LUNA collaboration (Laboratory for Underground Nuclear Astrophysics) has performed a high precision experiment on the $^3$He($\alpha,\gamma$)$^7$Be reaction in Italy's Gran Sasso underground laboratory (LNGS). Owing to the extremely low background conditions of the underground laboratory, the cross section has been measured at low energies never reached before. The details of the experiment have been discussed elsewhere \cite{bemmerer06,gyurky07,confortola07}. In the first phase of the experiment the cross section was measured with the activation technique at energies of 106, 127, 148 and 169 keV. In the second phase the same setup was used but the cross section was measured simultaneously with both on-line and activation methods. In this phase an energy point of 93\,keV has been measured and the points at 106 and 169 keV have been repeated. Table \ref{tab:results} summarizes the activation results obtained in the two phases of the LUNA experiments.

\section{Zero energy extrapolation of the LUNA activation data}

The high precision data obtained in the LUNA activation experiment can also be used to derive an extrapolated zero energy $S(0)$ factor. The following theoretical models are considered here: the resonating group calculation by Kajino and Arima \cite{kajino84}, the extended two-cluster model calculation by Cs\'ot\'o and Langanke \cite{csoto00} and the R-matrix calculation by Descouvemont \textit{et al.} \cite{desc04}. A $\chi^2$ minimalization fit has been carried out to the measured data where the absolute normalization of the theoretical data has been the only adjustable parameter. The theoretical cross section values at the energies where the LUNA measurements have been carried out were calculated in the following manner: In the case of Kajino and Arima \cite{kajino84} the energy dependence given by the formula (6) in Ref. \cite{kajino87} was adopted. For the model of Cs\'ot\'o and Langanke \cite{csoto00} linear energy dependence of the $S$ factor was supposed in the considered low energy range adopting the slope of the $S$ factor curve given in Table 1. of Ref. \cite{csoto00} using the MHN interaction. In the case of the R-matrix calculation of Descouvemont \textit{et al.} \cite{desc04} the $S$ factor values have been calculated with a linear interpolation between the two tabulated points closest to the measured energies. The low energy $S$ factor parametrization proposed by the NACRE compilation \cite{nacre} was also used to fit the data again leaving the absolute normalization as the only free parameter. The results of the fits are shown in Table \ref{tab:fit}.

\begin{table}
\caption{\label{tab:fit} $\chi^2$ fit to the LUNA activation data with different theoretical models and the NACRE parametrization}
\begin{tabular}{lcc}
\hline
model & $S(0)$ [keV b]& $\chi^2_{red}$ \\
\hline
Kajino and Arima \cite{kajino84} & 0.547 $\pm$ 0.017& 1.12 \\
Cs\'ot\'o and Langanke \cite{csoto00} & 0.586 $\pm$ 0.018& 1.47 \\ 
Descouvemont \textit{et al.} \cite{desc04} & 0.550 $\pm$ 0.017& 1.11 \\
NACRE \cite{nacre} & 0.597 $\pm$ 0.019 & 2.02 \\
\hline
\hline
\end{tabular}
\end{table}

\begin{figure}
\resizebox{\columnwidth}{!}{\rotatebox{270}{\includegraphics{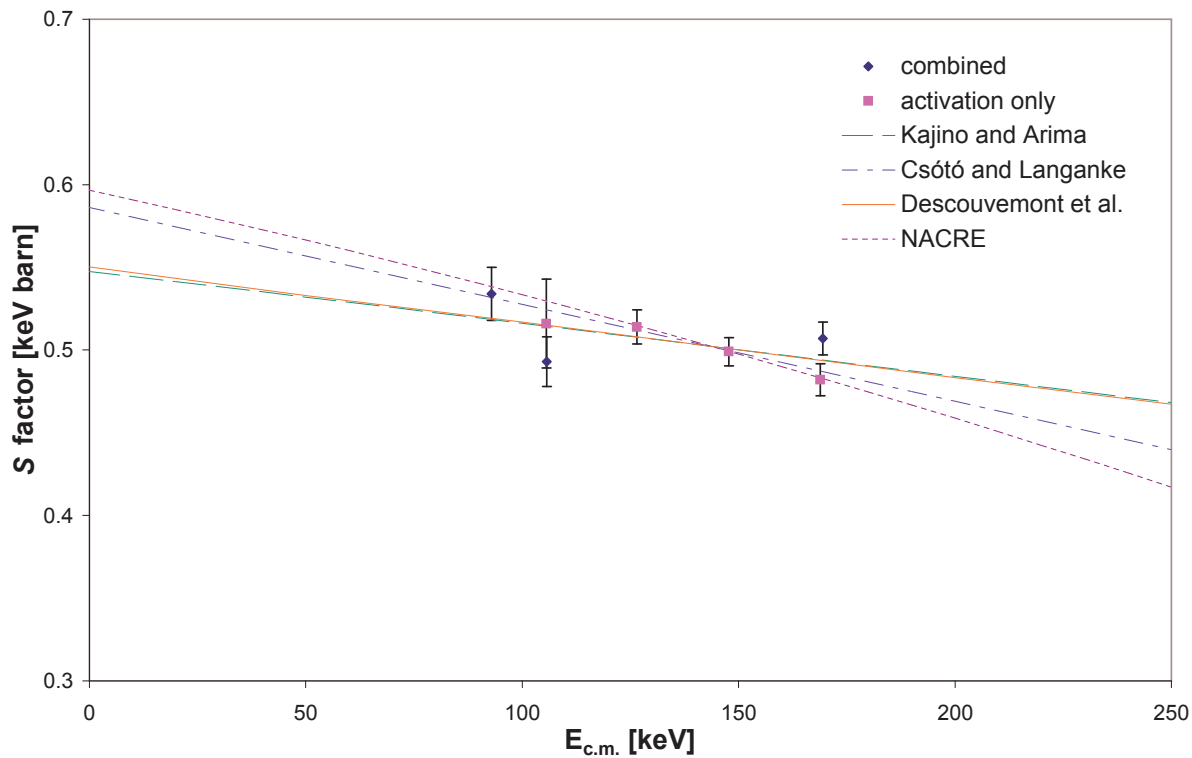}}}
\caption{Activation cross section of $^3$He($\alpha,\gamma$)$^7$Be measured by the LUNA collaboration. The calculated $S$ factors using different theoretical models normalized to the measured data are also plotted.}
\label{fig:luna_fit}
\end{figure}

The values of the reduced $\chi^2$ values have been calculated taking into account only the statistical uncertainties of the measured points. For the quoted uncertainty of the $S(0)$ values, 3\% systematic uncertainty was added quadratically to the error from the $\chi^2$ fit. As one can see, the models of Kajino and Arima \cite{kajino84} and Descouvemont \textit{et al.} \cite{desc04} give very similar results for both the reduced $\chi^2$ value and $S(0)$. The model of Cs\'ot\'o and Langanke \cite{csoto00} leads to a larger $\chi^2_{red}$ value and a considerably higher $S(0)$. The comparison of the measured data with the NACRE parametrization shows that the slope of the $S$ factor curve at low energies is underestimated by NACRE which leads to a high $S(0)$. This comparison is also shown in Fig.\,\ref{fig:luna_fit} where along with the measured data, the normalized theoretical curves from the above discussed theories and the NACRE parametrization are shown.

\section{Comparison of the results of various activation experiments}

The results of the R matrix calculations by Descouvemont \textit{et al.} \cite{desc04} are available in tabular form in a wide energy range. This provides a possibility for the comparison of the results of different $^3$He($\alpha,\gamma$)$^7$Be activation experiments. This can be done by renormalizing the R matrix curve to a certain set of data and deriving $S(0)$. Table~\ref{tab:actcomp} shows the resulting $S(0)$ values for the different activation experiments. The measurement of Volk \textit{et al.} \cite{volk83} is omitted since in that experiment only the energy integrated cross section has been determined. As one can see, there is a perfect agreement between the $S(0)$ values obtained by the two recent, high precision experiments carried out by Nara Singh \textit{et al.} and the LUNA collaboration. The experiment by  Robertson \textit{et al.} \cite{robertson83} leads to a significantly higher $S(0)$. This experiment was carried out at a single energy point only and may have had an additional systematic error. The work of Osborne \textit{et al.} \cite{osborne82} again results in a somewhat higher $S(0)$. Although this is statistically still consistent with the two most recent experiments, the higher obtained cross section values may indicate a hidden systematic error. Therefore, we suggest that for the zero energy $S$ factor from the activation experiments, the average of the two most recent results should be used leading to a value of $S(0)$\,=\,0.550\,$\pm$\,0.013.

\begin{table}
\caption{\label{tab:actcomp} Extrapolated $S(0)$ values from different activation experiments renormalizing the R matrix curve of Descouvemont \textit{et al.} \cite{desc04}}
\begin{tabular}{lc}
\hline
experiment & $S(0)$ [keV b] \\
\hline
Osborne \textit{et al.} \cite{osborne82} & 0.60 $\pm$ 0.04 \\
Robertson \textit{et al.} \cite{robertson83} & 0.67 $\pm$ 0.04 \\
Nara Singh \textit{et al.} \cite{narasingh04} & 0.551 $\pm$ 0.022 \\
LUNA activation \cite{bemmerer06,gyurky07,confortola07}& 0.550 $\pm$ 0.017 \\
\hline
\end{tabular}
\end{table}

\begin{figure}
\resizebox{\columnwidth}{!}{\rotatebox{270}{\includegraphics{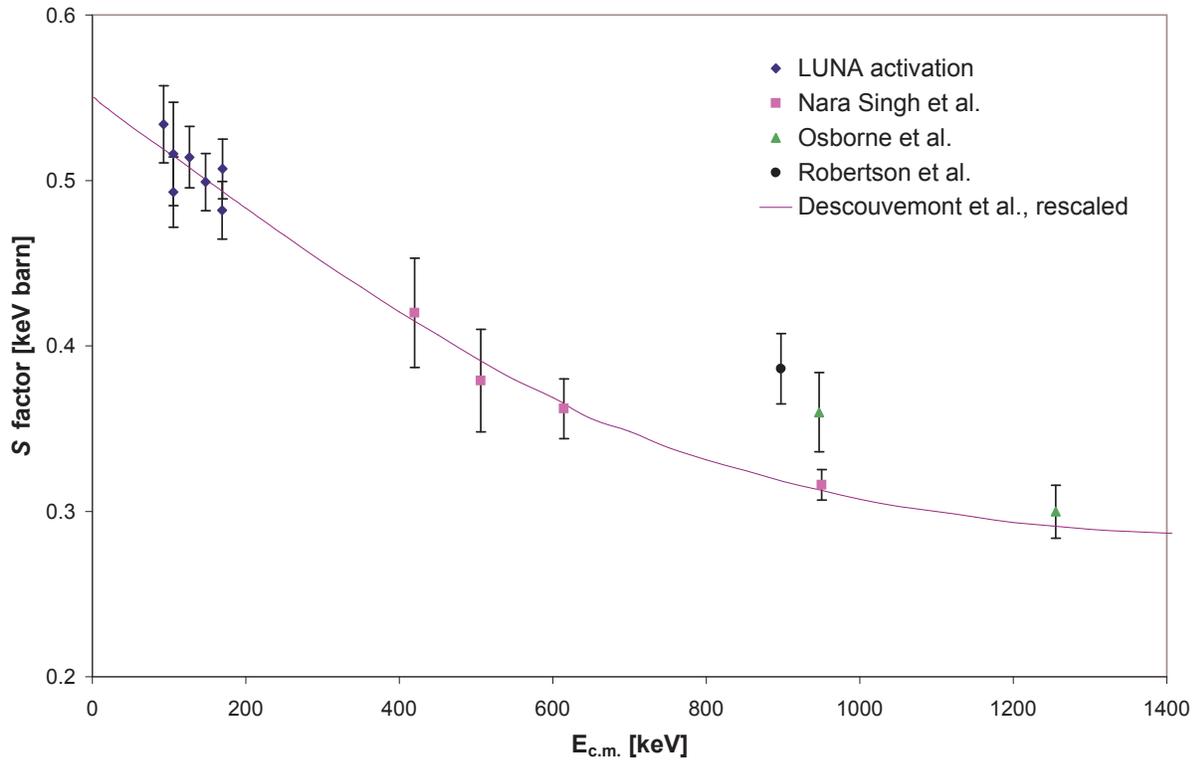}}}
\caption{The results of all activation experiments on the $^3$He($\alpha,\gamma$)$^7$Be reaction. The R-matrix calculation of Descouvemont et al. \cite{desc04} normalized to the LUNA and Nara Singh \textit{et al.} data is also shown.}
\label{fig:all_act_fit}
\end{figure}

Figure\,\ref{fig:all_act_fit} shows the measured $S$ factor data from the four experiments considered above. The R-matrix calculation by Descouvemont \textit{et al.} rescaled to the average of the LUNA and Nara Singh \textit{et al.} data is also plotted. The perfect agreement of the two recent experiments and the systematically higher values obtained in the older measurements is clearly visible.

The next step should be the inclusion of the on-line cross section data in the $S(0)$ analysis. The high precision LUNA on-line data are already available \cite{confortola07} but the analysis of these data with the comparison of other on-line experiments and the activation results is beyond the scope of the present paper.

\section*{Acknowledgments}

This work was supported by INFN and in part by the European Union (TARI RII3-CT-2004-506222), the Hungarian Scientific Research Fund (K68801 and T49245), and the German Federal Ministry of Education and Research (05CL1PC1-1).


\section*{References}
\begin{thebibliography}{10}
\bibitem{aharmim05} Aharmim B\textit{et al.} 2005 Phys. Rev. C \textbf{72} 055502
\bibitem{hosaka06} Hosaka J\textit{et al.} 2006 Phys. Rev. D \textbf{73} 112001
\bibitem{alimonti02} Alimonti G \textit{et al.} 2002  Astropart. Phys. \textbf{16} 205 
\bibitem{ahmad01} Ahmad Q R \textit{et al.} 2001 Phys. Rev. Lett. \textbf{87} 71301
\bibitem{eguchi03} Eguchi K \textit{et al.} 2003 Phys. Rev. Lett. \textbf{90} 21802
\bibitem{adelberger98} Adelberger E \textit{et al.} 1998 Rev. Mod. Phys. \textbf{70} 1265
\bibitem{narasingh04} Nara Singh B S, Hass M, Nir-El Y and Haquin G 2004 Phys. Rev. Lett. \textbf{93} 262503
\bibitem{bahcall04} Bahcall J N and Pinsonneault M H 2004 Phys. Rev. Lett. \textbf{92} 121301
\bibitem{spergel03} Spergel D \textit{et al.} 2003 Astrophys. J. Suppl. Ser. \textbf{148} 175
\bibitem{holmgren59} Holmgren H D and Johnston R L 1959 Phys. Rev. \textbf{113} 1556
\bibitem{parker63} Parker P D and Kavanagh R W 1963 Phys. Rev. \textbf{131} 2578
\bibitem{nagatani69} Nagatani K, Dwarakanath M R and Ashery D 1969 Nucl. Phys. A \textbf{128} 325
\bibitem{krawinkel82} Kr\"awinkel H \textit{et al.} 1982 Z. Phys. A \textbf{304} 307
\bibitem{osborne82} Osborne J L \textit{et al.} 1982 Phys. Rev. Lett. \textbf{48} 1664
\bibitem{alexander84} Alexander T K , Ball G C, Lennard W N, Geissel H and
Mak H B 1984 Nucl. Phys. A \textbf{427} 526
\bibitem{hilgemeier88} Hilgemeier M, Becker H W, Rolfs C, Trautvetter H P and Hammer J W 1988 Z. Phys. A \textbf{329} 243
\bibitem{robertson83} Robertson R G H \textit{et al.} 1983 Phys. Rev. C \textbf{27} 11
\bibitem{volk83} Volk H, Kr\"awinkel H, Santo R and Wallek L 1983 Z. Phys. A \textbf{310} 9l
\bibitem{bemmerer06} Bemmerer D {\it et~al.} 2006 Phys.~Rev.~Lett.  {\textbf 97} 122502 
\bibitem{gyurky07} Gy\"urky Gy \textit{et al.} 2007 Phys. Rev. C \textbf{75} 035805
\bibitem{confortola07} Confortola F \textit{et al.} 2007 Phys. Rev. C \textbf{75} 065803
\bibitem{kajino84} Kajino T and Arima A 1984 Phys. Rev. Lett. \textbf{52} 739
\bibitem{csoto00} Cs\'ot\'o A and Langanke K 2000 Few-Body Systems \textbf{29} 121
\bibitem{desc04} Descouvemont P \textit{et al.} 2004 Atomic Data and Nuclear Data Tables \textbf{88} 203
\bibitem{kajino87} Kajino T, Toki H, Austin S M 1987 Astrophys. J. \textbf{319} 531 
\bibitem{nacre} Angulo C \textit{et al.} 1999 Nucl. Phys. A \textbf{656} 3
\end{thebibliography}
\end{document}